# Prediction of $(TiO_2)_x(Cu_2O)_y$ Alloys for Photoelectrochemical Water Splitting

Heng-Rui Liu[1], Ji-Hui Yang[1], Yue-Yu Zhang[1], Shiyou Chen[2], Aron Walsh[3], Hongjun Xiang[*1], Xingao Gong[*1], and Su-Huai Wei[*4]

[1] *Key Laboratory of Computational Physical Sciences (Ministry of Education), State Key Laboratory of Surface Physics, and Department of Physics, Fudan University, Shanghai 200433, P. R. China*
[2] *Key Laboratory of Polar Materials and Devices (MOE), East China Normal University, Shanghai 200241, P. R. China*
[3] *Centre for Sustainable Chemical Technologies and Department of Chemistry, University of Bath, Claverton Down, Bath BA2 7AY, United Kingdom*
[4] *National Renewable Energy Laboratory, Golden, Colorado 80401, USA*

**ABSTRACT**

The formation of $(TiO_2)_x(Cu_2O)_y$ solid-solutions are investigated using a global optimization evolutionary algorithm. First-principles calculations based on density functional theory are then used to gain insight into the electronic properties of these alloys. We find that: (i) Ti and Cu in $(TiO_2)_x(Cu_2O)_y$ alloys have similar local environments as in bulk $TiO_2$ and $Cu_2O$ except for $(TiO_2)(Cu_2O)$ which has some trigonal-planar Cu ions. (ii) The predicted optical band gaps are around 2.1 eV (590 nm), thus having much better performance for the absorption of visible light compared with both binary oxides. (iii) $(TiO_2)_2(Cu_2O)$ has the lowest formation energy amongst all studied alloys and the positions of its band edges are found to be suitable for solar-driven water splitting applications.

A growing amount of attention is being placed on harvesting energy from renewable sources in order to meet the worldwide energy demand and relieve the global warming caused by our dependence on fossil fuels. In addition of direct conversion of sunlight to electricity via the photovoltaic effect, the production of a storable chemical fuel can be achieved via a photochemical process. Producing hydrogen from photoelectrochemical (PEC) water splitting is one of the most promising solutions[1,2] currently under extensive study.

Titanium dioxide, $TiO_2$, is an important photocatalyst due to its good charge transfer properties and its high resistance to photo- and chemical corrosion. It had been demonstrated that $TiO_2$ can decompose water into oxygen and hydrogen without the application of an external voltage[3] under certain conditions in the pioneering work of Fujishima and Honda four decades ago. Unfortunately, the common phases of $TiO_2$ have wide band gaps (3.2 eV for anatase phase and 3.0 eV for rutile phase), which can only be activated effectively under UV-radiation, but not by visible light. There is a great deal of literature [4-11] describing attempts to improve the sensitization of $TiO_2$ while maintaining its beneficial photocatalytic properties. The most common approaches involve doping $TiO_2$ by transition metal cations at Ti sites and/or anions at O sites.[6-8] Recently, structural modification has also been reported[9-11] to be an effective way to enhance the optical absorption of $TiO_2$ in the visible region of the electromagnetic spectrum.

Cuprous oxide, $Cu_2O$, which has been well studied for its semiconducting properties, has a fundamental band gap of 2.2 eV (563 nm), is also a promising material for the conversion of solar energy into electrical or chemical energy. Recently, it has been explored as a photocatalyst for solar-driven water splitting applications: a light to hydrogen conversion efficiency up to 26% at 400 nm was achieved for a $Cu_2O$ film with a [111] crystallographic orientation.[12-14] The main drawback for its usage as a photocathode is its weak light adsorption near the band edge (dipole forbidden band gap[15]) and instability in aqueous solutions.

It is natural to consider the combination of the two binary materials to form ternary $TiO_2/Cu_2O$ alloys, which may benefit from the attractive band gap of $Cu_2O$ as well as the high stability in aqueous solutions of $TiO_2$. The lower symmetry of the alloy should improve optical absorption near the band edge. Moreover, the addition of low binding energy Cu $d^{10}$ orbitals is beneficial, as the valence band (VB) edge of $TiO_2$ is much lower than the oxygen evolution potential, whereas the conduction band (CB) edge of $TiO_2$ is only slightly higher than the hydrogen evolution potential.[16] As a matter of fact, some experimental attempts have been made in this direction and various $TiO_2/Cu_2O$ composites, which can absorb in the visible region, have been synthesized.[17-22] However, due to the complexity of the resulting crystal structures, a detailed understanding of the structural and electronic properties of these $TiO_2/Cu_2O$ alloys is still lacking. In this paper, we investigate the stable structures of $(TiO_2)_x(Cu_2O)_y$ solid-solutions with different compositions using a global optimization evolutionary algorithm, and examine their electronic properties by first-principles calculations. Our findings suggest that these alloys, especially $(TiO_2)_2(Cu_2O)$, are strong candidates for visible-light-driven PEC water splitting applications because of their good optical absorption as well as their suitable band edge potentials.

We employed the global optimization methodology implemented in the *Universal Structure Predictor: Evolutionary Xtallography* (USPEX)[23,24] package for the investigation of the structures of $(TiO_2)_x(Cu_2O)_y$ alloys. We considered all possible cell shapes and sizes, with no symmetry constraints, for a maximum of 20 atoms per crystallographic unit cell. The total energies were calculated by density functional theory (DFT) using the semi-local Perdew-Burke-Ernzerhof (PBE) functional.[25] In the DFT plane-wave calculations, the ion-electron interaction was treated by the projector augmented wave (PAW) technique as implemented in the *Vienna ab initio simulation package* (VASP).[26,27] The non-local Heyd-Scuseria-Ernzerhof hybrid functional HSE06[28] was also employed to calculate the structural and electronic properties of the $(TiO_2)_x(Cu_2O)_y$ alloys obtained in order to give more quantitative results for the band structures. The Γ-centered Monkhorst-Pack k-point meshes[29] were used to achieve converged results for Brillouin zone integration. The phonon spectra were calculated with density functional perturbation theory (DFPT)[30] within the *Quantum Espresso* code.[31]

The predicted $(TiO_2)_x(Cu_2O)_y$ alloys at different proportions of $TiO_2$ and $Cu_2O$ are illustrated in Figure 1. The coordination numbers of Ti atoms and Cu atoms are the same with those in bulk $TiO_2$ and bulk $Cu_2O$ for almost all studied structures, that is, 6 (octahedral) for Ti and 2 (linear) for Cu. The main exception is the $(TiO_2)(Cu_2O)$ structure, where in addition to linear Cu, some trigonal planar Cu atoms are formed in order to satisfy the alloy stoichiometry (see the blue triangles in Figure 1). The bond lengths and angles of Ti-O and Cu-O bonds in the ternary alloys are also similar to those found in the parent binary oxides.

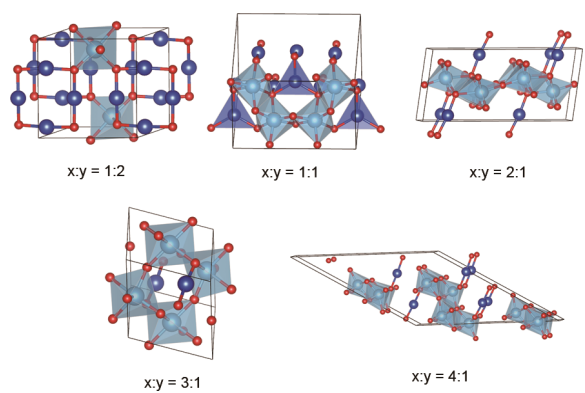

**Fig. 1** Structures of $(TiO_2)_x(Cu_2O)_y$ alloys, with different proportions of $TiO_2$ and $Cu_2O$. The small red balls are O, the blue ones are Cu, while the Ti ions are located at the center of light blue octahedra. The structure models are drawn using *VESTA*.[35]

The formation energies of $(TiO_2)_x(Cu_2O)_y$ alloys relative to anatase-structured $TiO_2$ and cuprite-structured $Cu_2O$ are shown in Figure 2(a). The formation energies are typically around a few tens of meV per atom and the curve shows an 'M' shape with the local minimum at x:y=2:1. First-principles molecular dynamics (MD) simulations have been carried out to confirm the thermal stability of these structures using the constant temperature and volume ensemble (timestep 1 fs) at 300 K. As $(TiO_2)_2(Cu_2O)$ has the lowest formation energy and is thus the most plausible phase among all studied alloys, we will focus mainly on its properties.

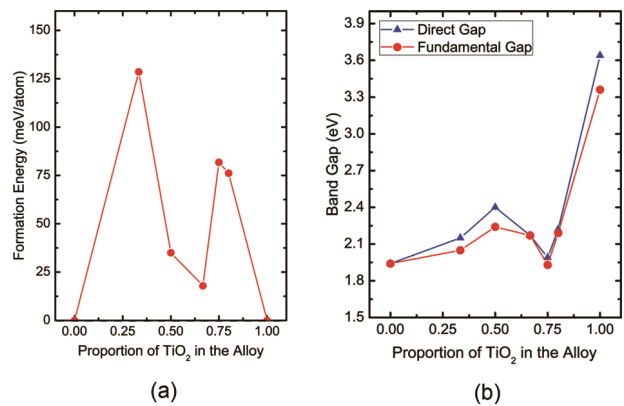

(a)             (b)

**Fig. 2** (a) Formation energy of $(TiO_2)_x(Cu_2O)_y$ relative to the binary oxides using the PBE functional, and plotted as a function of $x/(x+y)$. (b) Band gaps of $(TiO_2)_x(Cu_2O)_y$ alloys using the HSE06 functional.

The HSE06 band gaps of these alloys, along with $TiO_2$ and $Cu_2O$, are plotted in Figure 2(b). The band gap of $Cu_2O$ is 1.94 eV, slightly underestimated from the experimental value of 2.2 eV, while a slight overestimation of $TiO_2$ band gap is observed. Since the functional can reproduce the band gap of both bulk $Cu_2O$ and bulk $TiO_2$ in good agreement with experiments,[12,32,33] it is reasonable to assume that the predicted band gaps of the $(TiO_2)_x(Cu_2O)_y$ alloys are also reliable.

$(TiO_2)_2(Cu_2O)$ is found to be a direct gap material with a band gap of 2.17 eV (571 nm). The rest of the alloys have indirect band gaps, with direct gaps close in energy, of around 2.1 eV (590 nm), which makes them suitable for solar energy applications.

The stability of $(TiO_2)_2(Cu_2O)$ was further tested by an MD simulation at 600 K using a 486 atoms supercell. The results are shown in Figure 3(a), which demonstrate that the phase remains intact after 12 ns. We have also calculated its phonon structure [Figure 3(b)] and the absence of any soft mode confirms that the alloy is dynamically stable.

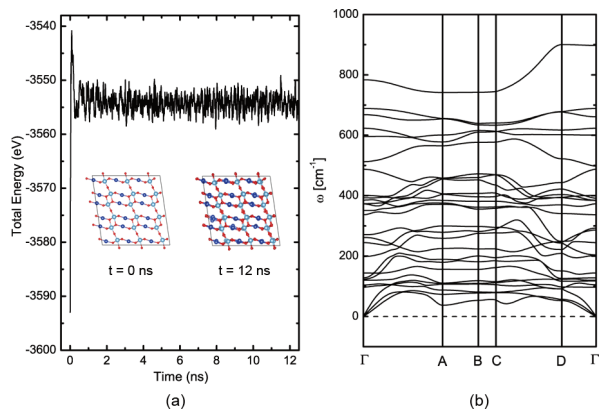

**Fig. 3** (a) Molecular dynamics simulation of $(TiO_2)_2(Cu_2O)$, and (b) its phonon band structure along a series of high symmetry lines.

The band structure and partial density of states (PDOS) of $(TiO_2)_2(Cu_2O)$, using the HSE06 functional, are plotted in Figure 4. The lowest conduction band and topmost valence band are rather flat, leading to a substantial density of states, which will help its optical absorption in the visible light region. Moreover, the upper valence bands are mainly derived from the hybridization between the O 2p and Cu 3d states, while the lower conduction bands are dominated by the empty 3d states of Ti. The atomic orbital components of the band edges of other alloys are similar to those of the $(TiO_2)_2(Cu_2O)$ (see ESI[†]).

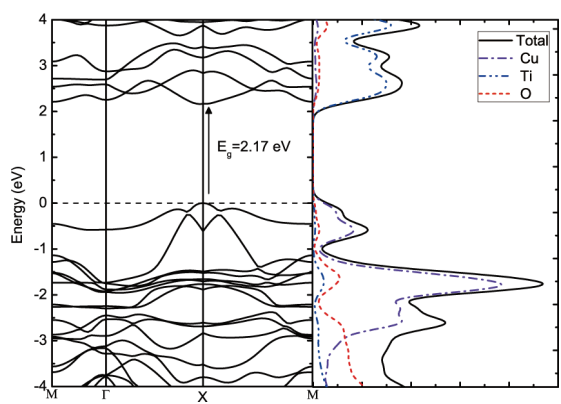

**Fig. 4** Band structure and partial density of states (PDOS) of the $(TiO_2)_2(Cu_2O)$ alloy calculated using the HSE06 functional. The PDOS is broadened by a Gaussian smearing of 0.2 eV.

The calculated optical absorption spectra of the $(TiO_2)_x(Cu_2O)_y$ alloys as well as the binary oxides are plotted in Figure 5. Anatase $TiO_2$ shows no absorption below 3.0 eV due to its wide band gap. $Cu_2O$ also has a low absorption coefficient in the 2.0 ~ 3.0 eV region, suffering from the forbidden transition at its fundamental band gap.[15] However, it can be clearly seen that all of the three alloys have much stronger absorption of visible light in the visible region compared to both bulk $Cu_2O$ and $TiO_2$. Indeed, $(TiO_2)_3(Cu_2O)$ exhibits the best absorption performance as a benefit from its small optical gap. The performance of $(TiO2)_2(Cu_2O)$ is between that of the $(TiO_2)(Cu_2O)$ and $(TiO_2)_3(Cu_2O)$ alloys.

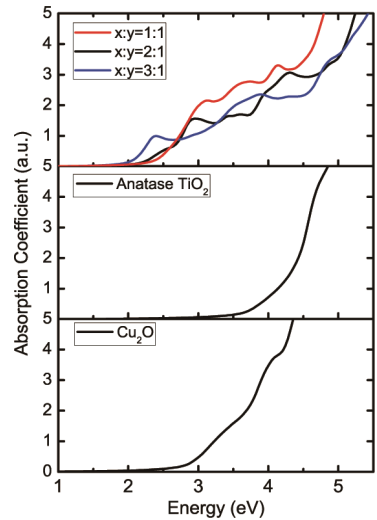

**Fig. 5** Predicted optical absorption of the $(TiO_2)_x(Cu_2O)_y$ alloys, $TiO_2$ and $Cu_2O$, using the HSE06 functional.

We have estimated the band edge potentials, referenced to the vacuum level,[34] of the $(TiO_2)_2(Cu_2O)$ alloy and anatase $TiO_2$, as illustrated in Figure 6. Our results for anatase $TiO_2$ show that its VB is 7.4 eV below the vacuum level, which is in good agreement with experiment (~7.5 eV).[16] As expected, a large VB offset between the alloy and anatase $TiO_2$ is observed, which mainly results from the low binding energy of the Cu 3d orbitals. On the other hand, due to quantum confinement of the isolated Ti-O layers (see Figure 1), the CB of the alloy is above that of anatase $TiO_2$, which indicates that the alloy could be more efficient than anatase $TiO_2$ in photocatalysis due to enhanced electron transfer rates. Consequently, the band alignment between anatase $TiO_2$ and the alloy is of type II. This suggests that at a $(TiO_2)_2(Cu_2O)/TiO_2$ heterojunction, the photogenerated holes and electrons can be separated and then accumulated at the alloy side and anatase TiO2 side, respectively. In Figure 6 we also marked the position of the hydrogen evolution potential and the position of oxygen evolution potential.[16] It is clear that the CB of the alloy is above the hydrogen evolution potential and the VB is below the oxygen evolution potential, which is suitable for PEC water splitting.

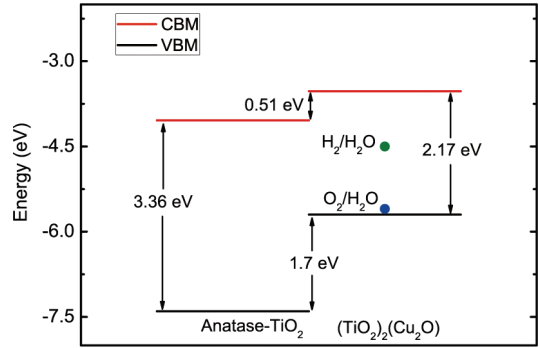

**Fig. 6** Band edges of the $(TiO_2)_2(Cu_2O)$ alloy and anatase $TiO_2$ relative to the vacuum level. The green dot represents the hydrogen evolution potential, while the blue one represents the oxygen evolution potential.

In conclusion, we show that (i) stable crystal structures of $(TiO_2)_x(Cu_2O)_y$ alloys can be obtained using the global optimization evolutionary algorithm; (ii) the optical band gaps of the obtained alloys are around 2.1 eV, and they absorb much more visible light than the $TiO_2$ and $Cu_2O$ component oxides; (iii) $(TiO_2)_2(Cu_2O)$ has the lowest formation energy among all studied alloys and its band edge energies are predicted to be suitable for visible-light-driven water splitting.

This work is partially supported by the Special Funds for Major State Basic Research, National Science Foundation of China, The Program for Professor of Special Appointment at Shanghai Institutions of Higher Learning, FANEDD, Ministry of Education and Shanghai Municipality. The calculations were performed in the Supercomputer Center of Fudan University. The work at NREL was supported by the U.S. Department of Energy under Contract No. DE-AC36-08GO28308. A.W. is supported by the Royal Society University Research Fellowship scheme.